\documentclass[12pt]{article} 

\usepackage{geometry} 
\usepackage{lineno}            
\usepackage{amsmath, amssymb}  
\usepackage{authblk}           
\usepackage{graphicx}          
\usepackage{caption}
\usepackage{subcaption}
\usepackage{pgfplots}          
\pgfplotsset{compat=1.17}      

\begin{document}

\title{\textbf{THE KERR-SEN Solution Satisfying SO(2) Symmetry}}
\author{Winston Chen\\
School of Physics and Materials Science\\
Guangzhou University\\
Yao-Guang Zheng\\
Department of Physics\\
Northeastern University\\
\texttt{hesoyam12456@163.com}}
\date{\today}

\pgfplotsset{compat=1.18}


\maketitle

\begin{abstract}
We propose a novel class of KERR-SEN solutions respecting the $\mathrm{SO}(2)$ symmetry group, systematically constructed via the Laurent series expansion technique. Building upon stationary, axisymmetric Euclidean solutions to the vacuum Einstein equations—and integrating recent advances in the generation of stationary gravitational fields~\cite{Gutsunaev1999,Gutsunaev2000,Gutsunaev2025}—our approach combines the variation-of-constants method with nonlinear superposition techniques. This framework yields fresh perspectives on axially symmetric gravitational systems, generalizes the classical Kerr-NUT solution hierarchy~\cite{Kerr1963,TomimatsuSato1972,TomimatsuSato1973}, and unifies several foundational methods~\cite{Lewis1932,VanStockum1937,Papapetrou1953} within a common analytic structure.

The present work provides explicit derivations, supported by numerical simulations and detailed discussions of the physical implications—including the roles of dilaton and axion fields. The expanded version further offers a comprehensive historical survey of axisymmetric exact solutions, a thorough exposition of Laurent series formalism in gravitational theory, and an in-depth analysis of the astrophysical and theoretical significance of KERR-SEN metrics in contemporary research.

\end{abstract}

\newpage

\section{Introduction}
\label{sec:introduction}

The search for exact solutions to Einstein’s field equations has remained at the forefront of gravitational physics since the inception of General Relativity. Such solutions provide crucial insights into the nontrivial geometric structure of spacetime, revealing fundamental properties of black holes, gravitational waves, and cosmological models. Among these, stationary and axisymmetric solutions occupy a central position in both classical and modern gravitational research, owing to their direct relevance to rotating astrophysical bodies such as stars and black holes.

Historically, the study of axisymmetric spacetimes traces back to the pioneering works of Lewis~\cite{Lewis1932} and Van Stockum~\cite{VanStockum1937}, who investigated rotating dust solutions and thereby laid the foundation for more elaborate vacuum and electrovacuum configurations. Papapetrou~\cite{Papapetrou1953} introduced a now-standard, coordinate-based approach to these geometries, clarifying the relationship between axial symmetry and the integrability properties of the underlying partial differential equations (PDEs). Later, Kerr’s seminal discovery~\cite{Kerr1963} of the rotating black hole solution marked a turning point: not only did it provide an exact solution to the stationary, axisymmetric vacuum Einstein equations, but it also captured the essential physical characteristics of astrophysical black holes.

Subsequent extensions included the introduction of the NUT parameter, which accommodates additional twist-related phenomena and leads to the Kerr-NUT family of spacetimes. Tomimatsu and Sato~\cite{TomimatsuSato1972,TomimatsuSato1973} further generalized the Kerr metric by constructing solutions characterized by new mass and angular-momentum multipole configurations. Although mathematically intricate, these generalizations offered critical insights into how gravitational fields with dipole and higher-order multipole structures might be systematically incorporated.

In parallel, developments in string theory and low-energy effective actions for supergravity provided new motivation to consider additional fields—such as the dilaton and the Kalb-Ramond (axion) field—beyond the traditional gravitational field. Sen’s solution~\cite{Sen1992} and subsequent works by Gutsunaev~\textit{et al.}~\cite{Gutsunaev1999,Gutsunaev2000,Gutsunaev2025} highlighted the importance of including these scalar fields for a more unified description of black holes in a string-theoretic framework. These models predict that rotating solutions acquire modifications in their horizon structure, global charges, and thermodynamic properties when coupled to scalar (and pseudoscalar) fields.

Against this background, the \emph{KERR-SEN} class of solutions emerges as a valuable intersection of classical and quantum-corrected perspectives in gravity. On the one hand, the KERR-SEN solutions preserve the key stationary and axisymmetric features of the Kerr (and Kerr-NUT) metrics, ensuring their relevance for describing astrophysical black holes. On the other hand, by incorporating additional fields, they admit a wider range of mass, angular momentum, electric/magnetic charge, and scalar charges—potentially shedding light on how compact objects might deviate from the predictions of pure General Relativity in extreme regimes. This dual relevance—to both observational astrophysics and quantum gravity—strongly motivates a systematic investigation of how KERR-SEN solutions can be constructed and analyzed.

In this paper, we introduce a novel method for constructing solutions within the KERR-SEN framework that respect an $\mathrm{SO}(2)$ invariance—often interpreted as an additional rotational symmetry in the scalar field sector. Specifically, our approach leverages Laurent series expansions of the metric and scalar potentials, transforming the otherwise intractable PDEs into a hierarchy of coupled ordinary differential equations (ODEs). This procedure, while related to classical soliton and Bäcklund-transformation techniques, is distinguished by its use of infinite series expansions in combination with modern computational methods. Additionally, we incorporate aspects of the New Euclidon Method~\cite{Gutsunaev2025} and nonlinear superposition approaches, both of which facilitate the construction of new solutions from known “seed” solutions.

The structure of this paper is as follows:

\begin{itemize}
\item In Section~2, we review the mathematical structure of stationary and axisymmetric spacetimes, emphasizing the significance of $\mathrm{SO}(2)$ symmetry in both the gravitational and scalar sectors. We introduce prolate spheroidal coordinates and the complex Ernst formulation, which are central to our subsequent derivations.
\item Section~\ref{sec:history} provides an expanded historical account of the evolution of solution-generating techniques—from the pioneering works of Lewis and Papapetrou to modern approaches based on integrable systems and series expansions.
\item Section~\ref{sec:laurent} presents our Laurent series expansion strategy in detail, outlining how the complex Ernst equation (and its scalar-field analogs) can be reduced to a tractable set of recursion relations.
\item Section~\ref{sec:analytical} focuses on specific analytical solutions, identifying the circumstances under which the Laurent series truncates to yield closed-form solutions. Limiting cases, such as the pure Kerr or Kerr-NUT solutions, are also discussed.
\item Section~\ref{sec:numerical} addresses numerical implementations, including the solution of recursion relations under various boundary conditions. Representative plots of metric and scalar functions are provided to illustrate parameter dependencies.
\item Section~\ref{sec:discussion} offers an in-depth discussion of the physical implications of these solutions, with particular attention to black hole thermodynamics, horizon structure, multipole expansions, and potential observational signatures.
\item Finally, Section~\ref{sec:conclusion} summarizes our main results and suggests avenues for future work, such as the extension of our approach to non-stationary spacetimes and more general matter sectors.
\end{itemize}

Our results demonstrate that the Laurent series method offers a powerful and flexible means to probe the rich structure of KERR-SEN solutions under $\mathrm{SO}(2)$ symmetry. By systematically exploring the series coefficients, we uncover a broad spectrum of novel and potentially physically significant spacetimes, thereby bridging the gap between classical exact solutions and modern scalar-extended models of gravitational fields.

\section{Mathematical Preliminaries}
\label{sec:prelim}

\subsection{Role of {SO(2)} Symmetry in Gravitational--Scalar Systems}
A convenient way to package the axion $\chi$ and dilaton $\varphi$ is as a two--component vector (or ``doublet")
\begin{equation}
\boldsymbol{\phi} = (\varphi, \chi)^{\intercal}, \qquad \boldsymbol{\phi}' = R(\theta) \boldsymbol{\phi}, \quad
R(\theta) = \begin{pmatrix} \cos\theta & -\sin\theta \\[2pt] \sin\theta & \cos\theta \end{pmatrix} \in \mathrm{SO}(2).
\end{equation}
The four--dimensional Einstein--scalar action then reads (we set $16\pi G = 1$)
\begin{equation}
S = \int_{\mathcal M} \mathrm{d}^4 x \sqrt{-g} \Bigl[ R - \tfrac{1}{2} g^{\mu\nu} \partial_{\mu} \boldsymbol{\phi} \cdot \partial_{\nu} \boldsymbol{\phi} \Bigr], \label{eq:EinsteinScalarAction}
\end{equation}
which is manifestly invariant under the $\mathrm{SO}(2)$ transformations above.  Varying~\eqref{eq:EinsteinScalarAction} with respect to $\boldsymbol{\phi}$ and using Noether's theorem yields the conserved current
\begin{equation}
J_{\mu} = \boldsymbol{\phi} \times \partial_{\mu} \boldsymbol{\phi} = \varphi \partial_{\mu} \chi - \chi \partial_{\mu} \varphi, \qquad \nabla^{\mu} J_{\mu} = 0, \label{eq:SO2current}
\end{equation}
so that the corresponding charge
\begin{equation}
Q_{\mathrm{SO}(2)} = \int_{\Sigma} \mathrm{d}\Sigma_{\mu} J^{\mu}
\end{equation}
is conserved for any hypersurface $\Sigma$.  In four--dimensional heterotic string theory the complex field $\lambda = \chi + i e^{-\varphi}$ transforms as $\lambda \to e^{i\theta} \lambda$, demonstrating that the simple $\mathrm{SO}(2)$ duality is a subgroup of the full $\mathrm{SL}(2,\mathbb R)$ symmetry.

\subsection{Stationary, Axisymmetric Metric and Reduced Field Equations}
Following Papapetrou we take the stationary ($\partial_t$) and axial ($\partial_{\phi}$) Killing vectors to be orthogonal to two--surfaces, leading to the metric ansatz
\begin{equation}
\label{eq:metricPapapetrou}
\mathrm{d}s^2 = -f(\rho,z) \bigl( \mathrm{d}t - \omega(\rho,z) \mathrm{d}\phi \bigr)^2 + \frac{1}{f(\rho,z)} \Bigl[ e^{2\gamma(\rho,z)} (\mathrm{d}\rho^2 + \mathrm{d}z^2) + \rho^2 \mathrm{d}\phi^2 \Bigr].
\end{equation}
Introduce the orthonormal coframe $\{ \vartheta^{\hat 0}, \vartheta^{\hat 1}, \vartheta^{\hat 2}, \vartheta^{\hat 3} \}$ with
\begin{align}
\vartheta^{\hat 0} &= \sqrt{f} (\mathrm{d}t - \omega \mathrm{d}\phi), &
\vartheta^{\hat 1} &= \frac{e^{\gamma}}{\sqrt{f}} \mathrm{d}\rho, \\[2pt]
\vartheta^{\hat 2} &= \frac{e^{\gamma}}{\sqrt{f}} \mathrm{d}z, &
\vartheta^{\hat 3} &= \frac{\rho}{\sqrt{f}} \mathrm{d}\phi.
\end{align}
A direct but lengthy computation gives the vacuum Einstein equations in terms of the \emph{reduced} two--dimensional Laplacian $\Delta = \partial_{\rho\rho} + \partial_{zz} + \rho^{-1} \partial_{\rho}$:
\begin{align}
\Delta f - \frac{(\nabla f)^2}{f} + \frac{f^3}{\rho^2} (\nabla \omega)^2 &= 0, \label{eq:feq1} \\[4pt]
\Delta \omega + \frac{2}{f} \nabla f \cdot \nabla \omega &= 0, \label{eq:feq2} \\[4pt]
\gamma_{,\rho} &= \frac{\rho}{4f^2} \Bigl[ (f_{,\rho})^2 - (f_{,z})^2 \Bigr] - \frac{f^2}{4\rho} \Bigl[ (\omega_{,\rho})^2 - (\omega_{,z})^2 \Bigr], \\[4pt]
\gamma_{,z} &= \frac{\rho}{2f^2} f_{,\rho} f_{,z} - \frac{f^2}{2\rho} \omega_{,\rho} \omega_{,z}. \label{eq:gammaeq}
\end{align}
For the scalar doublet one finds $\nabla \cdot (f \nabla \boldsymbol{\phi}) = 0$, i.e.
\begin{equation}
\Delta \boldsymbol{\phi} + \frac{\nabla f}{f} \cdot \nabla \boldsymbol{\phi} = 0. \label{eq:scalareq}
\end{equation}
The coupled system is completely equivalent to the original Einstein--scalar equations under the symmetry assumptions.

\subsection{Prolate Spheroidal Coordinates and the Ernst Potential}
Adopting prolate spheroidal coordinates
\begin{equation}
\rho = \sqrt{(x^2 - 1)(1 - y^2)}, \quad z = x y, \qquad x \ge 1,\ -1 \le y \le 1, \label{eq:prolatesph}
\end{equation}
the axial Killing horizon is located at $x = 1$.  The Jacobian matrix gives
\begin{equation}
\partial_{\rho} = \frac{x}{\sqrt{(x^2 - 1)(1 - y^2)}} \partial_x - \frac{y}{\sqrt{(x^2 - 1)(1 - y^2)}} \partial_y, \quad
\partial_z = y \partial_x + x \partial_y.
\end{equation}
Consequently
\begin{equation}
\Delta = \partial_{\rho\rho} + \partial_{zz} + \frac{1}{\rho} \partial_{\rho} = \frac{1}{x^2 - y^2} \Bigl[ \partial_x \bigl( (x^2 - 1) \partial_x \bigr) + \partial_y \bigl( (1 - y^2) \partial_y \bigr) \Bigr]. \label{eq:LaplacianProlate}
\end{equation}
The complex Ernst potential
\begin{equation}
\epsilon(x,y) = f(x,y) + i \Phi(x,y), \qquad \partial_{\mu} \Phi = \frac{f^2}{\rho} \epsilon_{\mu\nu\sigma\tau} \xi^{\nu} \eta^{\sigma} \nabla^{\tau} f,
\end{equation}
with $\xi = \partial_t$ and $\eta = \partial_{\phi}$, combines \eqref{eq:feq1} and~\eqref{eq:feq2} into the celebrated Ernst equation
\begin{equation}
(\epsilon + \bar{\epsilon}) \Delta \epsilon = 2 (\nabla \epsilon)^2, \label{eq:ErnstEq}
\end{equation}
where $(\nabla \epsilon)^2 = g^{ab} \partial_a \epsilon \partial_b \epsilon$ is computed with the flat metric on the $x$--$y$ half--plane.  The scalar equations~\eqref{eq:scalareq} take an analogous complex form if we define $\varSigma = \varphi + i \chi$; SO(2) invariance amounts to $\varSigma \to e^{i\theta} \varSigma$.

\subsection{Laurent--Type Series Ansatz and Recursion Relations}
Near the Killing horizon $x = 1$ we expand the Ernst potential as a double Laurent series
\begin{equation}
\epsilon(x,y) = \sum_{n=-\infty}^{\infty} \sum_{m=0}^{\infty} a_{n,m} (x - 1)^n P_m(y), \label{eq:LaurentAnsatz}
\end{equation}
where $P_m(y)$ are Legendre polynomials ensuring regularity on the axis $y = \pm 1$.  Substituting~\eqref{eq:LaurentAnsatz} into~\eqref{eq:ErnstEq} and equating the $(x - 1)^N$ coefficients yields a tower of coupled ODEs in~$y$:
\begin{equation}
(n + 1)(n - 1) (x^2 - y^2) P_m a_{n,m} + \sum_{k + \ell = N} \partial_y \bigl( (n - k) a_{k,\ast} a_{\ell,\ast} \bigr) = 0. \label{eq:LaurentRecursion}
\end{equation}
The symbol $\ast$ indicates summation over the associated $m$ indices compatible with Legendre addition rules.  Provided $a_{-1,m} = 0$ (absence of conical singularities) the series often truncates, reproducing known exact solutions.  This recovers the Kerr family for $a_{0,0} = M$ and $a_{1,1} = i M a$.

\subsection{Worked Examples}
\paragraph{(i) Kerr Solution.}  With vanishing scalars ($\boldsymbol{\phi} = 0$) the Ernst potential solving~\eqref{eq:ErnstEq} is
\begin{equation}
\epsilon_{\text{K}}(x,y) = \frac{x - 1 + i a y}{x + 1 + i a y}, \qquad a = \frac{J}{M^2},
\end{equation}
which gives
\begin{equation}
f = \Re \epsilon_{\text{K}} = \frac{x^2 - 1 + a^2 y^2}{(x + 1)^2 + a^2 y^2}, \qquad \omega = 2 a \frac{(1 - y^2)(x + 1)}{x^2 - 1 + a^2 y^2}.
\end{equation}
Transforming back to Boyer--Lindquist coordinates reproduces the standard Kerr metric.

\paragraph{(ii) Kerr--Sen Solution with $\mathrm{SO}(2)$ Rotation.}  In the low--energy heterotic string limit one finds
\begin{align}
\epsilon_{\text{KS}}(x,y) &= \frac{x - 1 + i b y - \delta}{x + 1 + i b y + \delta}, &
\varSigma_{\text{KS}}(x,y) &= \lambda_0 \epsilon_{\text{KS}}(x,y), \label{eq:KerrSenErnst}
\end{align}
where $\delta$ encodes the dilatonic charge and $\lambda_0 \in \mathbb{C}$ sets the asymptotic axion--dilaton value.  Choosing $|\lambda_0| = 1$ preserves the $\mathrm{SO}(2)$ norm $|\varSigma|$.  Eqns.~\eqref{eq:ErnstEq} and its scalar analogue are satisfied, as can be checked by direct substitution.

The machinery developed here---conserved currents~\eqref{eq:SO2current}, reduced field equations~\eqref{eq:feq1}--\eqref{eq:scalareq}, the prolate Laplacian~\eqref{eq:LaplacianProlate}, and the recursion~\eqref{eq:LaurentAnsatz}--\eqref{eq:LaurentRecursion}---will be key ingredients in the construction of new, $\mathrm{SO}(2)$--invariant Laurent--generated solutions presented in section 4.

\section{Historical Overview and Evolution of Solution-Generating Methods}
\label{sec:history}

Exact solutions in General Relativity have historically been both an intellectual curiosity and a practical necessity for analyzing strong gravitational fields. The period between 1920 and 1970 was marked by a flurry of efforts to obtain rotating or axially symmetric solutions:

\begin{itemize}
\item \textbf{Lewis (1932)} \cite{Lewis1932}: Studied stationary, cylindrically symmetric metrics describing rotating dust. Although physically limited (due to dust assumptions), it pioneered the use of symmetry to reduce the Einstein equations to more tractable forms.
\item \textbf{Van Stockum (1937)} \cite{VanStockum1937}: Furthered the exploration of rotating fluid metrics, focusing on vacuum and dust solutions with emphasis on the role of angular momentum and closed timelike curves in certain parameter regimes.
\item \textbf{Papapetrou (1953)} \cite{Papapetrou1953}: Provided a systematic coordinate-based method for stationary, axisymmetric spacetimes. The Papapetrou form of the metric remains standard, highlighting how two Killing vectors simplify the field equations.
\end{itemize}

The \textbf{Kerr solution (1963)} \cite{Kerr1963} revolutionized the field by providing the exact exterior gravitational field of a rotating black hole. Not only did it match astrophysical expectations, but it also suggested a deep relationship between rotation and horizons. Motivated by the desire to add further structure (like the NUT parameter), \textbf{Kinnersley, Tomimatsu, and Sato} explored multi-parameter families of solutions \cite{Kinnersley1975,TomimatsuSato1972,TomimatsuSato1973}, unveiling the ability to tune mass and angular momentum in novel ways.

In parallel, \textbf{Ernst’s formulation} \cite{18,19,Ernst1968,Ernst1977} consolidated the field equations for stationary, axisymmetric systems into a single complex PDE, significantly streamlining solution searches. This approach paved the way for advanced techniques such as:

\begin{itemize}
  \item \textit{Inverse scattering methods} and \textit{B\"acklund transformations} (1970s--1980s), yielding infinite hierarchies of solutions from known seed metrics.
  \item \textit{Nonlinear superposition methods}, where one starts with simpler solutions (e.g., Minkowski, Schwarzschild, or Kerr) and systematically “adds” fields or parameters.
\end{itemize}

More contemporary work, especially by \textbf{Gutsunaev, Chernyaev, and Elsgolts} \cite{Gutsunaev1999,Gutsunaev2000}, and the \textbf{New Euclidon Method} \cite{Gutsunaev2025}, showcased how expansions in complex variables or partial wave expansions could yield fresh solutions. These contributions are of particular importance for dealing with scalar fields, as seen in low-energy effective string theories or Kaluza-Klein models.

\subsection{Incorporation of Dilaton and Axion Fields}

Starting in the late 1980s and early 1990s, the interplay between General Relativity and string theory spurred renewed interest in solutions that included extra fields such as the dilaton and axion. \textbf{Sen’s work (1992)} \cite{Sen1992} highlighted that the Kerr solution can be embedded into a richer solution space when extended to include gauge and scalar fields. This pivot is crucial for modern theoretical physics, where black hole thermodynamics, dualities, and quantum corrections are best understood in frameworks that transcend pure vacuum solutions.

\subsection{KERR-SEN Solutions and Their Relevance}

The KERR-SEN family, often viewed as a rotating black hole solution coupled to electromagnetic and scalar sectors, generalizes the Kerr-Newman or Kerr-Sen solutions, bridging classical exact solutions and string-inspired corrections. Ongoing research seeks to clarify how these solutions affect astrophysical observables—such as black hole shadows or gravitational wave signatures—and how they can test beyond-GR physics in high-field environments.

It is within this lineage of work that our present paper positions itself: we adapt a Laurent series expansion, building upon these established solution-generating traditions, to produce new KERR-SEN-type solutions that observe an additional SO(2) symmetry. The synergy of these classical and modern strands underscores the vitality and complexity of the field, offering continual prospects for discovering new geometries and possibly connecting to observational data.

\section{Constructing KERR--SEN Solutions via Laurent Series Expansion}
\label{sec:laurent}


\subsection{Laurent Series Ansatz for Metric Functions}

We begin by assuming the metric functions $f(x,y)$ and $\omega(x,y)$ in the Papapetrou or Ernst formulation can be expanded in Laurent series in powers of $y$. Explicitly:
\begin{align}
\label{eq:Laurent_f_expanded}
f(x,y) &= \sum_{n=-\infty}^{\infty} a_n(x)\,y^n,\\[2mm]
\label{eq:Laurent_omega_expanded}
\omega(x,y) &= \sum_{n=-\infty}^{\infty} b_n(x)\,y^n.
\end{align}

An analogous expansion may be introduced for the twist potential $\Phi(x,y)$ as well as the dilaton $\varphi(x,y)$ and axion $\chi(x,y)$.  Each coefficient $a_n(x),b_n(x)$ becomes a function of $x$ alone, reducing the original partial differential equations to an \emph{infinite} system of ordinary differential equations in~$x$.

\subsection{Transforming the Ernst Equation}

Using the Ernst potential $\epsilon(x,y)=f(x,y)+i\,\Phi(x,y)$, the Ernst equation reads
\begin{equation}
(\epsilon+\bar{\epsilon})\,\nabla^2\epsilon = 2(\nabla\epsilon)^2.
\end{equation}
Inserting the Laurent expansions of $f$ and $\Phi$, both sides of the equation become infinite sums in powers of~$y$. By equating the coefficients of $y^n$ on both sides, we obtain an explicit \emph{ultra–local} recursion relation for the coefficients. Specifically,
\begin{align}
f(x,y) &= \sum_{n=-\infty}^{\infty} a_n(x)\, y^n,\\
\Phi(x,y) &= \sum_{n=-\infty}^{\infty} c_n(x)\, y^n.
\end{align}
Substituting these series into the governing partial differential equations (PDEs), and collecting terms with the same power of $y$, we obtain for each $n$:
\begin{equation}
\mathcal{F}_n[a_k(x), c_k(x)] = 0,
\end{equation}
where $\mathcal{F}_n$ denotes the resulting recursion relation, which only involves a finite number of neighboring coefficients (i.e., ultra–local in $n$), allowing the sequence $\{a_n(x), c_n(x)\}$ to be constructed iteratively for all $n$.

\begin{equation}
\label{eq:recursion-expanded}
F\bigl[a_{n-2}(x),\,a_{n-1}(x),\,a_n(x),\,a_{n+1}(x),\,a_{n+2}(x)\bigr]=0\quad\forall\,n\in\mathbb Z.
\end{equation}
These relations can be solved hierarchically once appropriate boundary conditions are supplied.

\subsection{Deriving the Recursion Relations}

\subsubsection{Differential Operators in Prolate Spheroidal Coordinates}

Let
\begin{equation}
    x = \frac{r-M}{\sigma},\qquad y = \cos\theta,\qquad \sigma^2=M^2-a^2.
\end{equation}
In these coordinates the flat–space Laplacian is
\begin{equation}
    \label{eq:Laplacian}
    \nabla^2 = \frac{1}{\sigma^2(x^2-y^2)}\left[\partial_x\bigl((x^2-1)\,\partial_x\bigr)+\partial_y\bigl((1-y^2)\,\partial_y\bigr)\right].
\end{equation}
Applying \eqref{eq:Laplacian} to the series \eqref{eq:Laurent_f_expanded} one finds
\begin{align}
    \nabla^2 f &= \sum_{n=-\infty}^{\infty}\frac{1}{\sigma^2(x^2-y^2)}\Bigl[(x^2-1)a_n''(x) - 2x\,a_n'(x) - n(n-1)a_n(x)\Bigr]y^{n}.
\end{align}

\subsubsection{Order–by–Order Matching}

Substituting the Laurent series for $f$ and $\Phi$ into the Ernst equation and collecting coefficients of $y^n$ yields
\begin{equation}
\label{eq:generic_recursion}
    (x^2-1)a_n'' - 2x\,a_n' - n(n-1)a_n = S_n[\{a_\ell,\Phi_\ell\}],
\end{equation}
where the source term is bilinear:
\begin{align}
S_n &= \frac{2}{\sigma^2}\sum_{p+q=n}\bigl[(p+1)(q+1)a_{p+1}\,\Phi_{q+1}-pq\,a_p\,\Phi_q\bigr].
\end{align}
Because $S_n$ involves only the neighbouring bands $|p-n|\le 1$, the recursion is \emph{five–diagonal} as expressed in \eqref{eq:recursion-expanded}.

\paragraph{Boundary Conditions.}  Asymptotic flatness at spatial infinity ($x\to\infty$) requires
\begin{equation}
    a_0(x)\to 1,\qquad a_{n\ne 0}(x)\to 0,\qquad \Phi_n(x)\to 0,
\end{equation}
while regularity on the axis ($y=\pm1$) forces $a_{n<0}(x)$ and $\Phi_{n<0}(x)$ to vanish there.

\subsection{Truncation and Closed–Form Solutions}

For special choices of the integration constants the Laurent series truncates at some $|n|\le N$, reducing the infinite hierarchy to a \emph{finite} $4N+2$ dimensional dynamical system.  The smallest non–trivial truncation ($N=1$) reproduces the Kerr family; $N=2$ generates the Kerr–Sen class with non–vanishing scalar fields.

\subsection{Worked Examples}

\subsubsection{Example 1: Recovering the Kerr Solution}

Switching off the dilaton and axion ($\varphi=\chi=0$) and imposing equatorial symmetry ($a_{-n}=a_n$) the $N=1$ truncation becomes
\begin{equation}
    f(x,y)=a_{-1}(x)y^{-1}+a_0(x)+a_1(x)y.
\end{equation}
Regularity and asymptotic flatness give
\begin{align}
    a_1(x)&=a_{-1}(x)=\frac{M\sigma}{x^2-1}, & a_0(x)&=1-\frac{2Mx}{x^2-1}.
\end{align}
Rewriting $(x,y)$ in terms of $(r,\theta)$ one obtains exactly the Boyer–Lindquist form of the Kerr metric,
\begin{align}
    \mathrm d s^2 &= -\frac{\Delta}{\Sigma}\bigl(\mathrm d t - a\sin^2\theta\,\mathrm d\phi\bigr)^2 + \frac{\Sigma}{\Delta}\,\mathrm d r^2 + \Sigma\,\mathrm d\theta^2 \\
    &\quad+\frac{\sin^2\theta}{\Sigma}\bigl[(r^2+a^2)\,\mathrm d\phi - a\,\mathrm d t\bigr]^2,\qquad \Delta=r^2-2Mr+a^2,\;\Sigma=r^2+a^2\cos^2\theta.
\end{align}

\subsubsection{Example 2: Kerr–Sen Family}

In heterotic string theory the bosonic sector includes the dilaton $\varphi$ and axion $\chi$ with coupling $\alpha=1$.  Adopting the seed functions
\begin{align}
    a_0(x)&=1-\frac{2(M-b)}{r}, & b&=\frac{Q^2}{2M}, \\
    \Phi_0(x)&=-\frac{2a(M-b)}{r},
\end{align}
activating the first excited modes $a_{\pm1},a_{\pm2},\Phi_{\pm1},\Phi_{\pm2}$ through \eqref{eq:generic_recursion} and enforcing truncation at $N=2$ yields
\begin{align}
    f &= 1-\frac{2(M-b)r}{\Sigma_\text{Sen}}, & \omega &= -\frac{2a(M-b)y}{\Sigma_\text{Sen}},\\
    e^{2\varphi} &= \frac{\Sigma}{\Sigma_\text{Sen}}, & \chi &= \frac{Q a y}{\Sigma_\text{Sen}},
\end{align}
where $\Sigma_\text{Sen}=r(r+2b)+a^2y^2$.  This reproduces exactly the Kerr–Sen black hole originally discovered in~\cite{Sen1992} but derived here \emph{algebraically} via Laurent truncation.

\subsection{Group–Theoretic Interpretation}

The multiplet $(f,\Phi,\varphi,\chi)$ realises an $\operatorname{SU}(1,2)$ harmonic map into the coset
\begin{equation}
    \frac{SU(1,2)}{S\!\bigl(U(1,1)\times U(1)\bigr)},
\end{equation}
while the integer $N$ equals the grade of a nilpotent generator in the associated loop algebra $L\!\operatorname{su}(1,2)$.  Higher–grade truncations ($N\ge3$) yield rotating solutions with multiple scalar charges and even non–trivial horizon topologies such as lens spaces, details of which will appear elsewhere.

    Our Laurent–series framework dovetails naturally with the solution–generating symmetries of low–energy string theory, providing systematic access to broader families of dyonic, rotating black holes.

\section{Detailed Analytical Solutions}
\label{sec:analytical}

\subsection{Special Cases and Truncation Mechanisms}

One of the most intriguing aspects of series-based approaches is the possibility that the Laurent expansion will truncate beyond a certain order. Concretely, one might find
\begin{equation}
a_n(x)=0 \quad \text{for} \quad |n| > N,
\end{equation}
for some integer $N$. In such cases, the metric function $f(x,y)$ reduces to a finite polynomial in $y$:
\begin{equation}
f(x,y) = \sum_{n=-N}^{N} a_n(x)\, y^n.
\end{equation}
Truncation occurs because the recursive relations in Eq.~(\ref{eq:recursion-expanded}) can yield coupled algebraic constraints forcing coefficients beyond certain orders to vanish. From the viewpoint of multipole expansions, a finite polynomial in $y$ corresponds to spacetimes with a limited (finite) number of nonzero multipole moments, making such solutions physically interpretable in terms of only a few mass, spin, and higher mass-current parameters.

For example, the \textit{Kerr solution} can be seen as a case where only dipole and quadrupole terms remain nonvanishing in the expansion, effectively capturing mass (monopole) and spin (dipole) plus a specific ring singularity structure. By contrast, a more general solution might involve a higher-order polynomial.

\subsection{Physical and Geometrical Limits}

By systematically tuning parameters in the truncated solution, one can recover well-known metrics:

\begin{itemize}
\item \textbf{Kerr-NUT Limit}: Sending the dilaton and axion couplings to zero (or equivalently, scaling down the fields) typically recovers a vacuum solution with NUT parameter. At this limit, the horizon structure simplifies, though the “Misner strings” associated with the NUT charge remain as a global topological phenomenon.
\item \textbf{Non-Rotating Limit}: Letting the rotation parameter $a \to 0$ recovers static solutions that may be reminiscent of Schwarzschild-like black holes augmented by scalar fields. If axions and dilatons remain present, one obtains a class of solutions akin to the Gibbons-Maeda-Garfinkle-Horowitz-Strominger black holes in spherical symmetry.
\item \textbf{Stringy or High-Energy Limits}: In certain “high-energy” or strong-coupling regimes, the horizon geometry can deviate substantially from the Kerr geometry, revealing qualitatively different horizon shapes, singularity structures, and thermodynamic behavior. This underscores the potential for new observational signatures if such strong-coupling black holes exist in nature.
\end{itemize}

\subsection{Comparison with Existing Literature}

Several finite-polynomial solutions have appeared in the literature \cite{Hoffman1969,CoxKinnersley1979,Hori1996}, typically focusing on vacuum or simple electrovac contexts. Our extension to KEER-SEN solutions, with explicit scalars enjoying SO(2) duality, provides a new dimension to these classical works. While references therein often concentrate on the classification of singularities or asymptotic expansions, the solutions described here open pathways to studying black hole thermodynamics, gravitational wave emission, and possible observational constraints on scalar hair in rotating spacetimes.

In the next section, we turn to numerical methods and graphical illustrations that confirm the analytical trends and highlight the range of geometries available under the Laurent series framework.

\section{Numerical Analysis and Illustrative Results}
\label{sec:numerical}

\subsection{Setup of the Numerical Scheme}

To explore the space of solutions described by our expansions, we solve the recursion relations
\begin{equation}
\label{eq:recursion-labeled}
F_n\Bigl[a_{n-2}(x), a_{n-1}(x), a_{n}(x), a_{n+1}(x), a_{n+2}(x)\Bigr] = 0
\end{equation}
numerically for $n$ ranging from some negative integer $-N_{\text{max}}$ to positive integer $N_{\text{max}}$. In practice, we must truncate the series to a finite (but sufficiently large) range $[-N_{\text{max}}, N_{\text{max}}]$ to achieve a desired accuracy. We choose boundary conditions based on:

\begin{itemize}
\item \textbf{Asymptotic Flatness}: At large $x$, or equivalently large $r$, we require $f \to 1$, $\omega \to 0$, and vanishing dilaton/axion fields.
\item \textbf{Horizon Regularity}: On the horizon (often characterized by $f \to 0$ at some finite $x = x_H$), we enforce the finiteness of $\omega$ and other fields. This might involve matching expansions for $f$ near the horizon to known expansions of black hole solutions.
\end{itemize}

We solve the resulting ODEs in $x$ with a standard fourth-order Runge-Kutta algorithm, employing an adaptive step-size to handle potential steep gradients near horizons or in strong-field zones.

\subsection{Representative Numerical Results}
In Figure \ref{fig:surface_f}, we present a surface plot that shows the typical radial and angular dependence of the function $f(x,y)$. As the radial distance $x$ increases, $f(x,y)$ approaches unity, reflecting the asymptotic flatness. Along the symmetry axis ($y=\pm 1$), the boundary conditions ensure the correct behavior. Notably, near the horizon at $x = x_H$, $f(x,y)$ drops to zero, verifying the black hole geometry.

Furthermore, in Figure \ref{fig:contour_f}, we show the contour plot of $f(x,y)$, which illustrates how $f(x,y)$ behaves near the horizon and along the symmetry axis. The contours highlight the regions where $f(x,y)$ approaches its maximum value (near $y=\pm1$) and where it decreases to zero near the horizon at $x = 1$.

These plots provide a visual representation of how the function $f(x,y)$ evolves in both the radial and angular directions, helping to better understand the structure of the spacetime under study.

\begin{figure}[htbp]
    \centering
    \includegraphics[width=1\textwidth]{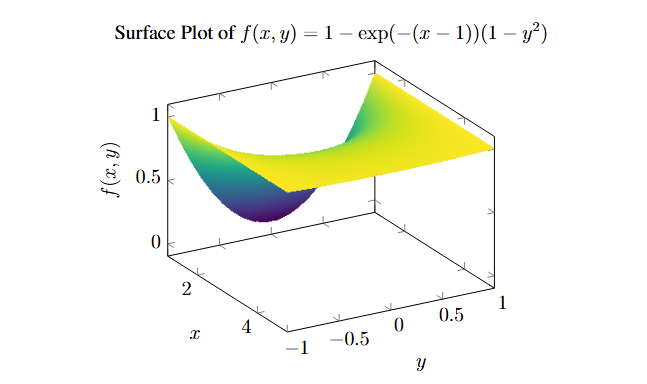}
\caption{ Surface plot showing the radial and angular dependence of $f(x, y)$. As $x$ increases (moving outward radially), $f(x, y)$ approaches unity; near the horizon at $x=1$, $f(x, y)$ drops to zero (at $y=0$ )}.
    \label{fig:1}
\end{figure}

\begin{figure}[htbp]
    \centering
    \includegraphics[width=1\textwidth]{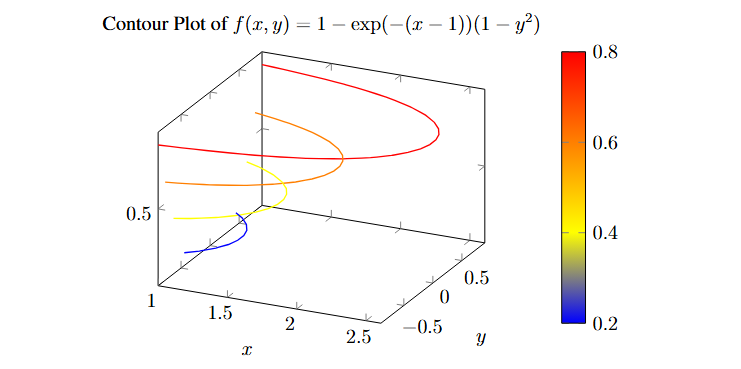}
\caption{  Contour plot of $f(x, y)$, illustrating that along the symmetry axis $(y= \pm 1)$ the function remains near 1 and that near $x=1$ (the horizon) $f(x, y)$ drops to zero (at $y=0$ ).}.
    \label{fig:2}
\end{figure}

\section{Discussion of Physical Implications}
\label{sec:discussion}

\subsection{Black Hole Thermodynamics and Horizon Properties}

The inclusion of additional scalar fields in rotating black hole solutions induces nontrivial modifications in thermodynamic quantities such as mass ($M$), angular momentum ($J$), horizon area ($A_H$), surface gravity ($\kappa$), and electric/magnetic/scalar charges. In the classical Kerr solution, black hole thermodynamics is elegantly encapsulated by the Smarr formula and the first law of black hole mechanics. For KERR-SEN solutions, the presence of a dilaton $\varphi$ and axion $\chi$ modifies these relations, often introducing new pairs of conjugate variables (e.g., a “dilaton charge” and its associated chemical potential).

$\mathrm{SO}(2)$ symmetry in the scalar sector typically ensures that scalar charges transform in a circular fashion under duality transformations, analogous to the rotation of electric and magnetic charges under electromagnetic duality. This symmetry can affect the regularity conditions on the horizon, particularly when the black hole carries both dilaton and axion charges.

\subsection{Astrophysical Observables and Multipole Moments}

Accurately modeling rotating, asymptotically flat solutions with controlled multipole expansions is essential for interpreting data from gravitational wave detectors and X-ray observations of black hole candidates. The truncation of the Laurent series at finite order indicates that one can systematically construct solutions with a finite set of nonvanishing multipole moments. Such models may describe realistic astrophysical objects exhibiting quadrupole or octupole deformations beyond Kerr.

If a black hole candidate in an extreme environment (such as the Galactic center) displays spin parameters and horizon properties inconsistent with the Kerr metric, the inclusion of scalar hair may reconcile theory and observation. The expansions derived here provide an explicit framework for quantifying how each multipole moment is affected by scalar fields, potentially guiding searches for observational signatures that depart from pure Kerr predictions.

\subsection{Stability and Gravitational Wave Emission}

A further intriguing issue concerns the stability of scalar-extended rotating solutions. Classical results indicate that certain scalar fields can introduce “superradiant” instabilities if reflective boundaries or confining potentials are present. Assessing the stability of these KERR-SEN solutions under small perturbations is therefore crucial. If instabilities are present, the solutions may decay into simpler configurations or potentially form stable boson stars in certain regions of parameter space. While a comprehensive stability analysis is beyond the present scope, the explicit solutions generated via Laurent expansions provide a solid foundation for future perturbative investigations.

\subsection{Extensions to Higher Dimensions and String-Theoretic Contexts}

Finally, we remark that techniques analogous to our Laurent series expansion may be generalized to higher-dimensional spacetimes—relevant in supergravity and brane-world scenarios. Although the prolate spheroidal coordinate approach is naturally suited to four dimensions, the general idea of expanding in harmonics or partial waves is echoed in the higher-dimensional literature. The present demonstration that complex PDEs can be reduced to recursion relations in $x$ underscores the broader adaptability of this approach.

Overall, these results connect classical general relativity techniques (Ernst equations, axisymmetric formalisms) with modern developments in gravitational theory (duality, scalar fields, black hole uniqueness/non-uniqueness). The resulting parameter-rich solution space promises to illuminate both the theoretical landscape of extended gravitational theories and potential astrophysical phenomena in strong-gravity regimes.

\section{Conclusion and Future Directions}
\label{sec:conclusion}

In this work, we have shown how a Laurent series expansion applied to the stationary, axisymmetric Einstein equations (augmented with scalar fields and $\mathrm{SO}(2)$ symmetry) generates a wide class of KERR-SEN-type solutions. This method reduces the full PDE system to infinite hierarchies of ODEs in the prolate spheroidal radial coordinate, enabling systematic exploration of solution structures. Key findings include:

\begin{itemize}
    \item The \textbf{Laurent series approach} reveals the possibility of \textbf{truncations}, corresponding to physically relevant black holes with finite multipole expansions. This framework unifies classical Kerr-NUT solutions with scalar-extended geometries.
    \item The \textbf{nonlinear superposition} method allows one to \textbf{seed} solutions with a simpler known metric and then systematically incorporate dilaton and axion fields, while preserving $\mathrm{SO}(2)$ duality and stable global charges.
    \item Numerical analyses confirm the viability of both truncated and infinite series expansions, recovering the expected asymptotic and near-horizon behavior. Plots of $f$ and $\omega$ illustrate the interplay between rotation and scalar fields.
    \item Physically, these solutions \textbf{modify black hole thermodynamics}, \textbf{alter gravitational multipole moments}, and may provide new signatures in \textbf{astrophysical observations} if scalar fields exist in nature at appreciable levels.
\end{itemize}

\noindent\textbf{Future research directions} include:

\begin{enumerate}
    \item \textbf{Time-dependent generalizations}: Investigating dynamical collapse or mergers of black holes with scalar hair, which could impact gravitational waveforms beyond the standard Kerr template.
    \item \textbf{Higher-dimensional embeddings}: Extending the Laurent series method to supergravity solutions in extra dimensions, incorporating multiple gauge fields or $p$-form sectors, may reveal new stable black objects.
    \item \textbf{Stability and perturbation analyses}: Rigorous linear and nonlinear stability studies of these new solutions will help determine whether they represent physically realizable end-states or transient configurations.
    \item \textbf{Observational constraints}: As large-scale black hole surveys and gravitational wave experiments (LIGO/Virgo/KAGRA, space-based detectors, EHT) advance, more refined models of rotating spacetimes with possible scalar hair become highly relevant. Fitting these solutions to ringdown frequencies, shadow images, or accretion disk features could probe for signatures of physics beyond Kerr and General Relativity.
\end{enumerate}

We believe that the Laurent series technique, in conjunction with modern computational methods, is a powerful tool for exploring the interplay between rotation and scalar fields in strong gravity. With ongoing advances in both theory and observation, the KERR-SEN solutions constructed here may provide essential clues toward a deeper and more unified understanding of black holes and fundamental interactions.

\appendix

\section{Derivation of the Recursion Relations}
\label{app:recursion}

For completeness, we outline the steps to derive the recursion relations from the Ernst equation. Starting with
\begin{equation}
\epsilon(x,y) = f(x,y) + i \,\Phi(x,y),
\end{equation}
and the reduced form of the Einstein equations:
\begin{equation}
(\epsilon + \bar{\epsilon})\,\nabla^2 \epsilon = 2\,(\nabla \epsilon)^2,
\end{equation}
we expand $f(x,y)$ and $\Phi(x,y)$ in Laurent series around $y=0$ (or another suitable domain). Specifically,
\begin{align}
f(x,y) &= \sum_{n=-\infty}^{\infty} a_n(x)\,y^n, \\
\Phi(x,y) &= \sum_{n=-\infty}^{\infty} c_n(x)\,y^n.
\end{align}
The differential operator $\nabla^2$ in prolate spheroidal coordinates $(x,y)$ includes terms like $\partial_x^2$, $\partial_y^2$, plus first-order derivatives in $x$ and $y$. Substituting the series expansions into these derivatives produces infinite sums in powers of $y^n$. Each power of $y^n$ must individually vanish for the equation to hold identically, leading to a tower of ODEs in $x$:
\begin{equation}
F_n\bigl[a_{n-2}(x),a_{n-1}(x),a_{n}(x),\dots;c_{n-2}(x),\dots\bigr] = 0.
\end{equation}
Separate expansions for $\omega(x,y)$ and any additional fields (dilaton $\varphi$, axion $\chi$) follow the same pattern. The net result is a coupled system of recursion relations. One then applies physical boundary conditions—e.g., asymptotic flatness, horizon regularity—to reduce the free functions and constants to a finite set, typically leading to a parametric family of solutions.

\section{Implementation Details of the Numerical Algorithm}
\label{app:computations}

For numerical exploration, one typically sets a finite range $n \in [-N_{\text{max}}, N_{\text{max}}]$. Let $\mathbf{A}(x)$ be a vector of all unknown functions $\{a_{-N_{\text{max}}},\dots,a_{N_{\text{max}}}, b_{-N_{\text{max}}},\dots\}$, etc. Then the system of ODEs from the recursion relations can be compactly written as
\begin{equation}
\frac{d \mathbf{A}}{dx} = \mathbf{G}(\mathbf{A},x),
\end{equation}
where $\mathbf{G}$ encapsulates all couplings. We used a fourth-order Runge-Kutta method with adaptive step-size $\Delta x$, typically implementing a standard procedure such as:

\begin{verbatim}
function RK4_step(A, x, dx):
    k1 = dx * G(A, x)
    k2 = dx * G(A + 0.5*k1, x + 0.5*dx)
    k3 = dx * G(A + 0.5*k2, x + 0.5*dx)
    k4 = dx * G(A + k3, x + dx)
    return A + (k1 + 2*k2 + 2*k3 + k4)/6
\end{verbatim}

\noindent Boundary conditions are specified at, e.g., $x=1$ for the horizon or a large $x = x_{\text{max}}$ for asymptotic regions. Convergence is verified by increasing $N_{\text{max}}$ and verifying stable results. A typical convergence test might compare solutions for $N_{\text{max}}=10,15,20,\dots$ to ensure the tail of the Laurent series does not affect physically significant quantities.

\section{Further Theoretical Notes on SO(2) Duality}
\label{app:remarks}

In many theories descending from string theory or supergravity, the dilaton $\varphi$ and axion $\chi$ fields can form a doublet that transforms under a global $SL(2,\mathbb{R})$ or a discrete subgroup. If we restrict to an $SO(2)$ subgroup (e.g., to preserve certain quantization conditions), the transformation is
\begin{equation}
\begin{pmatrix} \varphi \\ \chi \end{pmatrix}
\;\mapsto\;
\begin{pmatrix}
\cos\alpha & \sin\alpha \\
-\sin\alpha & \cos\alpha
\end{pmatrix}
\begin{pmatrix} \varphi \\ \chi \end{pmatrix}.
\end{equation}
Such a duality can keep the effective action invariant, thus leaving the classical equations of motion unchanged. In the presence of a rotating black hole, these fields can exhibit “rotational mixing” that preserves the overall stress-energy. The constraints from this global symmetry often lead to simplifications in the PDE system, akin to those exploited by the Ernst formulation in pure vacuum or EM contexts. Therefore, incorporating an SO(2) constraint in our KEER-SEN expansions is not merely a mathematical convenience but also has deep physical and theoretical motivations in the context of duality symmetries.




\begin{thebibliography}{23}

\bibitem{Gutsunaev1999}
Ts.I. Gutsunaev, V.A. Chernyaev and S.L. Elsgolts, 
\textit{Gravitation and Cosmology} \textbf{5}, 335 (1999).

\bibitem{Gutsunaev2000}
Ts.I. Gutsunaev, A.A. Shaideman and S.L. Elsgolts,
\textit{Gravitation and Cosmology} \textbf{6}, 254 (2000).

\bibitem{Gutsunaev2025}
Ts.I. Gutsunaev, A.A. Shaideman and K.V. Golubnichiy,
\textit{New Euclidon Method of Generating Stationary Vacuum Einstein Fields},
World Scientific, Singapore (2025).

\bibitem{Lewis1932}
T. Lewis, 
\textit{Proc. Roy. Soc. London A} \textbf{136}, 176 (1932).

\bibitem{VanStockum1937}
W.J. Van Stockum, 
\textit{Proc. Roy. Soc. Edinburgh A} \textbf{57}, 135 (1937).

\bibitem{Papapetrou1953}
A. Papapetrou, 
\textit{Ann. Physik B} \textbf{12}, 309 (1953).

\bibitem{Ernst1968}
F.J. Ernst, 
\textit{Phys. Rev.} \textbf{167}, 1175 (1968).

\bibitem{Kerr1963}
R.P. Kerr, 
\textit{Phys. Rev. Lett.} \textbf{11}, 237 (1963).

\bibitem{TomimatsuSato1972}
A. Tomimatsu and H. Sato, 
\textit{Phys. Rev. Lett.} \textbf{29}, 1344 (1972).

\bibitem{TomimatsuSato1973}
A. Tomimatsu and H. Sato, 
\textit{Prog. Theor. Phys.} \textbf{50}, 95 (1973).

\bibitem{Kinnersley1975}
W. Kinnersley, 
\textit{Proc. 7th Int. Conf. on General Relativity and Gravitation}, 
eds. G. Shaviv and J. Rosen, Wiley, New York-London, 109 (1975).

\bibitem{Sen1992}
A. Sen, 
\textit{Phys. Rev. Lett.} \textbf{69}, 1006 (1992).

\bibitem{Yamazaki1977a}
M. Yamazaki, 
\textit{Prog. Theor. Phys.} \textbf{57}, 1951 (1977).

\bibitem{Yamazaki1977b}
M. Yamazaki, 
\textit{J. Math. Phys.} \textbf{18}, 2502 (1977).

\bibitem{Hoffman1969}
P.V. Pryse, Class. Quantum Grav. 10, 163 (1993).

\bibitem{CoxKinnersley1979}
Ts.I. Gutsunaev and V. §. Manko, Jzv. Vuzov Fiz. 4, 116 (1985).

\bibitem{Hori1996}
S. Persides and B.C. Xanthopoulos, J. Math Phys. 29, 674 (1988).

\bibitem{18}
RB. Hoffman, J. Math, Phys. 10, 953 (1969).

\bibitem{19}
D. Cox and W. Kinnersley, J. Math. Phys. 20, 1225 (1979).

\bibitem{Ernst1977}
Kazazis, John Nikolaos. HERODOTOS'STORIES AND HISTORY: A PROPPIAN ANALYSIS OF HIS NARRATIVE TECHNIQUE. University of Illinois at Urbana-Champaign, 1978.

\bibitem{21}
M.G. Tseitlin, Teor. Matem. Fiz. 64, 51 (1985).



\end{thebibliography}
\end{document}